\documentclass[oldversion]{aa}
\usepackage{graphicx}
\begin{document}
\title{Near-infrared observations of water-ice in OH/IR stars
}
\author{K. Justtanont \inst{1} \and G. Olofsson \inst{1}
\and C. Dijkstra\inst{2} \and A.W. Meyer\inst{3}
}

\offprints{K. Justtanont}

\institute{Stockholm Observatory, AlbaNova, Dept. of Astronomy, 
SE-106 91 Stockholm, Sweden
\and Dept. Physics \& Astronomy, U. of Missouri Culombia, 300
Physics Building UMC, Columbia, MO 65211, USA
\and USRA/SOFIA, NASA Ames Research Center, MS 211-3, Moffett Field, 
CA  94035, USA
}

\date{Received ; accepted }

\abstract{
A search for the near-infrared water-ice absorption band was made in a number
of very red OH/IR stars which are known to exhibit the 10$\mu$m silicate
absorption. %using SIRCA. The camera has a wide and a narrow water-ice band
As a by-product, accurate
positions of these highly reddened objects are obtained. We derived a dust
mass loss rate for each object by modelling the spectral energy distribution
and the gas mass loss rate by solving the equation of motion for the
dust drag wind. The derived  mass loss rates show a strong correlation
with the silicate optical depth as well as that of the water-ice.
The stars have a high mass loss rate ($> 10^{-4}$ M$_{\odot}$ yr$^{-1}$)
with an average gas-to-dust mass ratio of 110.
In objects  which show the 3.1$\mu$m water-ice absorption, the near-IR slope
is much steeper than those with no water-ice. Comparison between our 
calculated mass loss rates and those derived from
OH and CO observations indicates that these stars have recently
increased their mass loss rates.
\keywords{Stars: AGB and post-AGB --
Stars: circumstellar matter -- Stars: evolution --  
Stars: late-type
-- Stars: mass-loss -- Infrared: stars}
}
\maketitle

\section{Introduction}

In order to estimate the dust mass loss rate in O-rich Asymptotic Giant Branch
(AGB) stars, it is common to use the strength of the 10$\mu$m silicate emission. 
There
seems to be a direct correlation between the 10$\mu$m excess emission and the
derived mass loss rate estimated by CO (Skinner \& Whitmore \cite{skinner}). However,
this does not apply when the silicate feature becomes self-absorbed or in absorption
as mass loss rate increases. On the contrary, it has been known that extreme 
OH/IR stars with strong silicate absorption show a very weak CO line strength
which indicates low derived gas mass loss rate (Heske et al. \cite{heske}).
Justtanont et al. (\cite{kay96}) investigated one such star, OH26.5+0.6  and
found that the discrepancy can be explained by the fact that the star has
recently undergone a superwind where the mass loss rate increases by two orders
of magnitude. The low rotational transitions of CO mainly trace the 
cool gas mass loss rate in the outer part of the envelope, which has not been affected
by this mass loss increase. The interferometric CO J=1-0 map using BIMA confirms
the envelope size is marginally resolved (Fong et al. \cite{fong}), consistent with
a compact CO envelope due to the superwind.

Accompanied by the deep 10$\mu$m silicate absorption, the near-infrared 
spectrum of some extreme OH/IR stars shows the absorption at  3.1$\mu$m due to 
water-ice. Meyer et al. (\cite{meyer}) suggested a correlation between the
mass loss rate and the column density of water-ice in a few sample of stars
observed. So far, there has not been a systematic search for the 3.1$\mu$m
absorption in these stars as a long integration is needed in order to
detect the absorption band when the infrared flux is already very low.
A medium resolution spectrum is needed in order to discriminate between
the type of ice condensed, i.e., amorphous or crystalline. Maldoni et al.
(\cite{maldoni}) suggested a significant fraction of water-ice in
crystalline form condensed in the outflow of OH32.8-0.3. The crystalline
ice has sharp features at 3.1, 44 and 62$\mu$m, the latter two were first 
detected by the observation by the Kuiper Airborne Observatory (Omont
et al. \cite{omont}). The libration band at 12$\mu$m is more difficult to
discern due to the fact that it blends with the strong silicate absorption.
Observations of OH/IR stars using the Infrared Space Observatory (ISO)
Short- and Long Wavelength Spectrometers
(SWS and LWS) show that these stars exhibit signatures of crystalline
silicate dust (e.g., Cami et al. \cite{cami}),
as well as water-ice (Sylvester et al. \cite{roger}).

\section{Observations}

\begin{table*}
\centering
\caption{Observed fluxes of the OH/IR stars using SIRCA.}
\begin{tabular}{lllcccc}
\hline\hline
Star    &  \multicolumn{2}{c}{J(2000)} & 
           \multicolumn{4}{c}{Flux (W m$^{-2} \mu$m$^{-1}$)}\\
\hline
        &  RA     &  Dec  &   K  &   L$^{\prime}$  & w-H$_{2}$O & n-H$_{2}$O \\
\hline
IRAS~18052-2016 & 18 08 16.38 & -20 16 11.5& 4.58E-14 & 1.10E-12 & 8.19E-13 & 7.20E-13\\
IRAS~18092-2347 & 18 12 16.97 & -23 47 06.1& 1.40E-13 & 4.58E-13 & 4.81E-13 & 4.27E-13\\
IRAS~18100-1915 & 18 13 03.07 & -19 14 19.0& 5.85E-17 & 3.36E-14 & 1.40E-14 & 2.49E-15$^{a}$\\
OH~16.1-0.3     & 18 21 07.11 & -15 03 24.5& 3.65E-13 & 3.73E-15$^{a}$ & 1.47E-12 & 1.41E-12\\
IRAS~18257-1052 & 18 28 30.96 & -10 50 48.5& 4.32E-11 & 1.13E-13 & 1.55E-13 & 1.38E-13\\
OH~21.5+0.5     & 18 28 30.94 & -09 58 14.6& 1.16E-16$^{a}$ & 5.36E-13 & 9.27E-14 & 2.64E-14\\
IRAS~18327-0715 & 18 35 29.22 & -07 13 10.9& 2.20E-14 & 2.90E-12 & 1.44E-12 & 1.20E-12\\
OH~26.5+0.6     & 18 37 32.51 & -05 23 59.2&  -  & 6.85E-11 & 9.42E-12 & 7.05E-12\\
IRAS~18361-0647 & 18 38 50.54 & -06 44 49.6& 1.49E-15 & 2.11E-13 & 8.26E-14 & 5.47E-14\\
OH~30.7+0.4     & 18 46 05.80 & -01 59 17.5& 2.48E-14 & 1.50E-12 & 6.49E-13 & 5.73E-13\\
OH~30.1-0.7     & 18 48 40.82 & -02 50 29.4& 2.46E-17 & 4.60E-13 & 9.80E-14 & 2.93E-14\\
IRAS~18488-0107 & 18 51 26.25 & -01 03 52.4& 3.31E-16 & 1.82E-13 & 4.22E-14 & 1.49E-14\\
OH~32.8-0.3     & 18 52 22.26 & -00 14 11.8& 3.42E-16 & 1.66E-12 & 2.30E-13 & 3.74E-14\\
IRAS~18588+0428 & 19 01 20.06 & +04 32 31.6& 6.74E-14 & 1.80E-12 & 1.36E-12 & 1.23E-12\\
OH~39.9-0.0     & 19 04 09.74 & +06 13 16.4& 1.28E-14 & 1.41E-12 & 6.84E-13 & 5.70E-13\\
IRAS~19067+0811 & 19 09 08.22 & +08 16 37.2& 1.76E-15 & 5.94E-14 & 1.43E-14 & 1.05E-14\\
IRAS~22177+5936 & 22 19 27.47 & +59 51 21.7& 1.67E-12 & 2.17E-11 & 2.04E-11 & 3.04E-11\\
\hline
$^{a}$ an upper limit flux
\end{tabular}
\label{tab-obs1}
\end{table*}

A sample of OH/IR stars were selected based on the deep 10$\mu$m silicate 
absorption, mainly from the {\it IRAS} low resolution spectra (LRS) and the
OH maser observations by Sevenster (\cite{seven}).
Our observations were done using the Stockholm InfraRed CAmera
(SIRCA) on the Nordic Optical Telescope (NOT) during July 22-23, 2003.
We observed a total of 17 stars in the nodding and chopping mode.
The stars were observed in the K, L$^{\prime}$, narrow ($\Delta \lambda = 0.15\mu$m)- 
and wide-H$_{2}$O-ice ($\Delta \lambda = 0.3\mu$m) bands. 
The central wavelength of the water-ice bands is at 3.07$\mu$m.
The field-of-view is relatively large, 1$^{\prime} 15^{\prime\prime}$
by 1$^{\prime} 15^{\prime\prime}$.
Each star is flat fielded then flux calibrated using standard
stars taken during the observation. A summary of the observations is given 
in Table~\ref{tab-obs1}.

Since we have images in several bands, it is a simple matter to identify
the OH/IR stars and determine their positions
with the aid of the 2MASS images. The images at
the K band in our observations are compared to the published map.
In some cases, a star is not visible in the K band so an interpolation
of the positions of nearby  stars was performed. Many of 
the fields show multiple point sources and with the aid of the L$^{\prime}$
band, most can be distinguished as the OH/IR stars. For the field of
star IRAS22177+5936, there is one very bright star which may be the
target as the flux increases from K to L$^{\prime}$ bands.
The measured flux in the 
narrow band is greater than the wide band at 3$\mu$m which is most likely
due to the saturation problem. We maintain this star in
our sample as we have CGS4 and archive ISO data.
From the difference of the wide- and narrow-water-ice band fluxes, we 
put the detection limit of the water-ice at 20\% of the difference over the
wide-band flux. 
We chose this limit based on the star OH26.5+0.6 which
was shown to have water-ice absorption at 3.1$\mu$m (Meyer et al. 
\cite{meyer}).

As a complement to our data set, earlier observations of six OH/IR stars
done at the 3.8-metre UKIRT, using the Cooled Grating Spectrometer (CGS4)
in August 1999 are included to enlarge the SIRCA sample. Some of the stars
overlap with the SIRCA sample (Table~\ref{tab-obs2}). For most stars, 
we have a spectral coverage between 2 to 4$\mu$m, punctuated by the earth's 
atmospheric absorption. The spectra were wavelength calibrated using arc spectra
and then flux calibrated with standard stars. The absolute flux level is
further adjusted to coincide with known photometry from either published 
data or SIRCA observations. This procedure is done simply because the
positions of these stars were not well known and the flux in the K band
is very low. Some stars are mispointed by several arcsec which means the
flux level can be much lower than expected. This can be seen by comparing
the coordinates of stars which are observed by both SIRCA and CGS4. 
Also, these stars are known to be variable with a period of up to 1\,000 days.

\begin{table}
\caption{A list of OH/IR stars using CGS4 and their observed positions.}
\begin{tabular}{lll}
\hline\hline
                & RA(2000)   & DEC(2000)   \\
\hline
AFGL~5379       & 17 44 23.9 & -31 55 39.4 \\
OH~13.1+5.0     & 17 55 45.1 & -15 03 43.1 \\
OH~16.1-0.3     & 18 21 06.9 & -15 03 20.9 \\
OH~21.5+0.5     & 18 28 31.8 & -09 58 30.0 \\
IRAS~18310-2834 & 18 34 13.7 & -28 32 21.4 \\
OH~30.1-0.7     & 18 48 42.0 & -02 50 36.5 \\
OH~39.9-0.0     & 19 04 09.8 & +06 13 16.3 \\
OH~51.8-0.2     & 19 27 41.9 & +16 37 23.3 \\
OH~65.7-0.8     & 19 59 39.1 & +28 23 07.2 \\
OH~75.3-1.8     & 20 29 08.5 & +35 45 43.8 \\
IRAS~22177+5936 & 22 19 27.3 & +59 51 22.3 \\
\hline
\end{tabular}
\label{tab-obs2}
\end{table}

\section{Determination of mass loss rates}

\begin{figure*}[t]
\centering
\includegraphics[width=17cm]{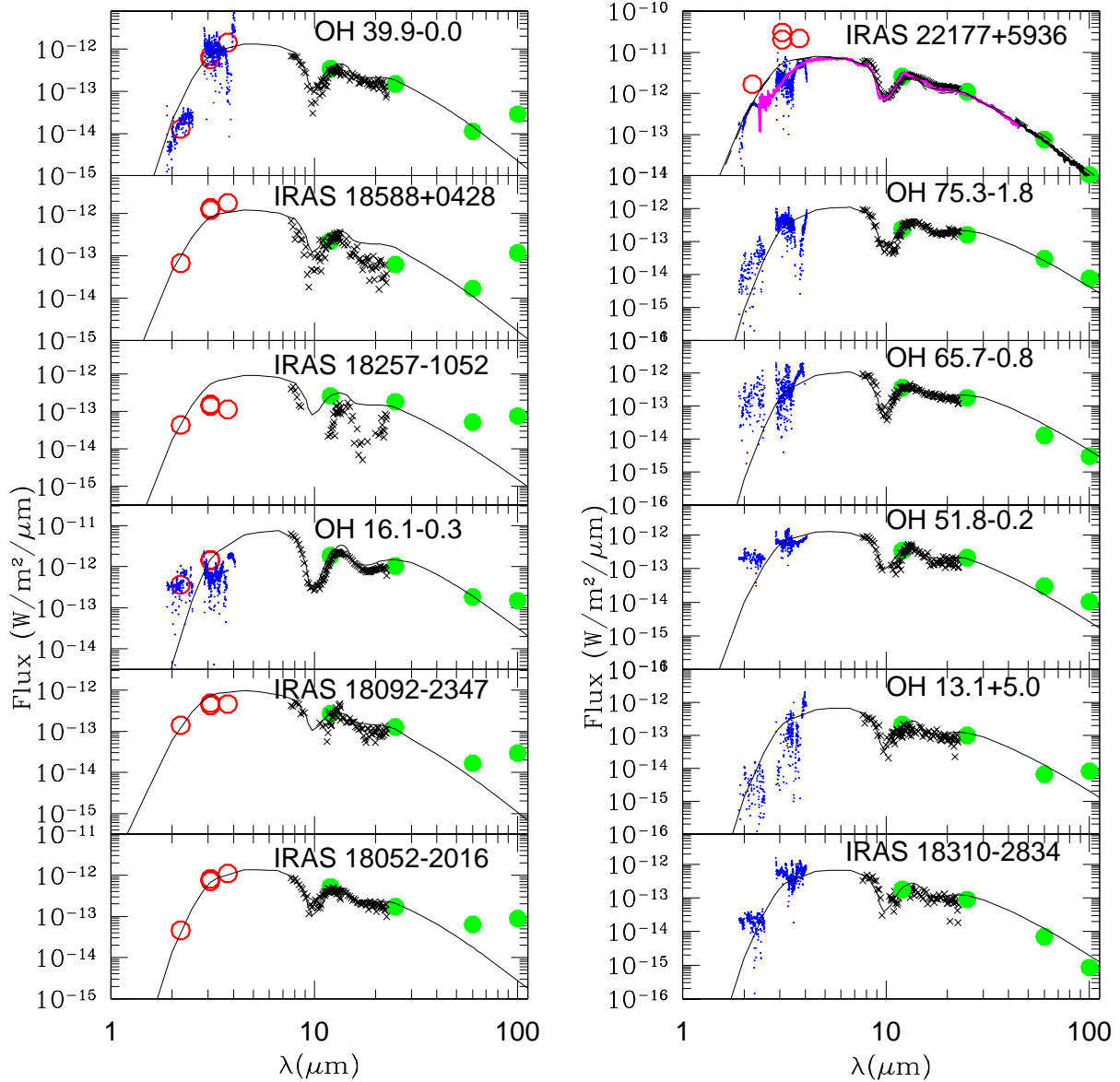}
\caption{SED fit to the stars which do not show ice absorption.
The symbols in the plots are x for the {\it IRAS} LRS, filled circles
are the {\it IRAS} photometry, open circles are the SIRCA photometry
and plus signs are CGS4 spectra. ISO spectra are shown as heavy solid line.
}
\label{fig-sed1}
\end{figure*}

In order to derive the dust mass loss rate from our sample stars, a
model fit to the spectral energy distribution (SED) of individual star
is performed. This, in turn, is used to calculate the total mass loss rate
assuming the momentum driven wind.

\begin{figure}[t]
\centering
\resizebox{\hsize}{!}{\includegraphics{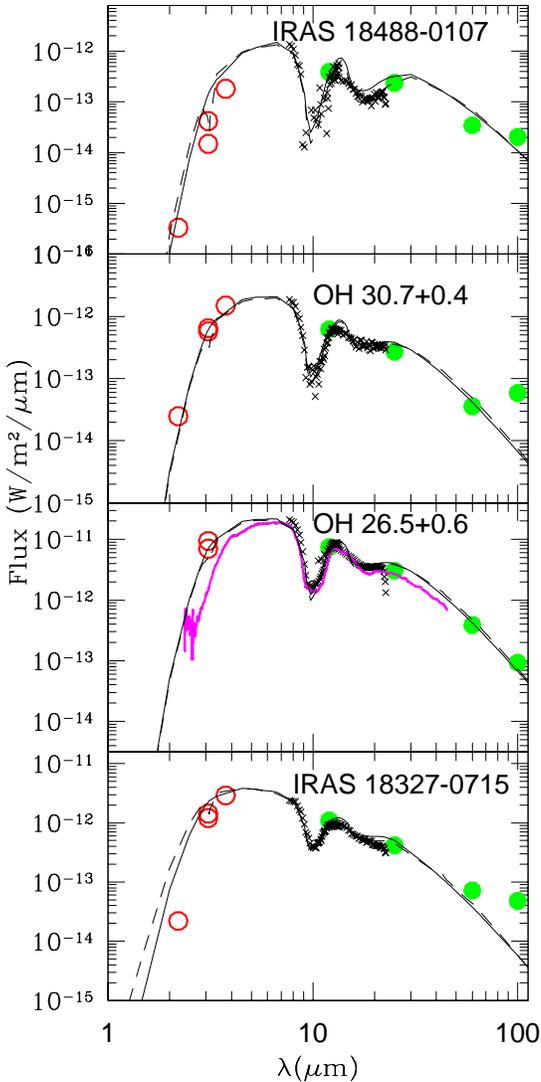}}
\caption{SED fit to the stars which show ice absorption.
The same symbols are used for photometric data and {\it IRAS} LRS.
The model with pure silicate component is shown as solid line while
the composite silicate core-ice mantle is shown as dashed line.
}
\label{fig-sed2}
\end{figure}

\subsection{Dust mass loss rates}

We derive the dust mass loss rate by fitting the SED for each star
(Justtanont \& Tielens \cite{kay92}).
We extracted {\it IRAS} LRS spectra which cover the wavelength range 
between 7 to 24$\mu$m, along with the photometry from the
point source catalogue. Where available, we also extracted archived 
ISO-SWS spectra. These
complementary data are crucial in the modelling as the major dust component 
is silicates which have absorption features in the 10$\mu$m window.

The radiative transfer calculation used to model the stars is based on the 
work by Haisch (\cite{haisch}). It assumes a spherically symmetric outflow, 
with a constant mass loss rate and outflow velocity. The thermal emission and 
scattering is calculated at a specified radial grid point and the emergent
flux and the temperature profile of the dust calculated. We assume a grain
size distribution for the dust to be the same as that of the interstellar
medium (Mathis et al. \cite{mrn}). As the distance to these stars are unknown, 
we assume a luminosity of 10$^{4}$L$_{\odot}$ and vary the distance until the 
overall spectrum fits.

We use two different grain models to fit our SED in the envelope which
shows water-ice absorption, namely, pure silicate grains and silicate core-water
ice mantle.
The dust parameters used to fit the absorption features are silicates based
on OH26.5+0.6 model by Justtanont et al. (\cite{kay96}) and crystalline
water-ice from Schmitt et al. (\cite{schmitt}). We select the crystalline
form of water-ice because there is evidence that the grains condensed in
such an outflow show sharp structure at 3.1$\mu$m (Maldoni et al. \cite{maldoni})
and on the bases of the far-IR features due to crystalline water-ice in
the ISO spectra (e.g., Sylvester et al. \cite{roger}; Dijkstra et al. \cite{rien1})
The inner radius of dust shell is set to a point where the temperature
is below the amorphous silicate condensation temperature of 1\,000~K while
the silicate core-ice mantle grains have a maximum temperature of 100~K,
i.e., the temperature which water-ice condenses onto existing
silicate grains.
The terminal velocity of the outflow of each star is well determined by
either CO rotational line or via OH maser observations (see Table~\ref{tab-mdot}).

\begin{figure*}[t]
\centering
\includegraphics[width=17cm]{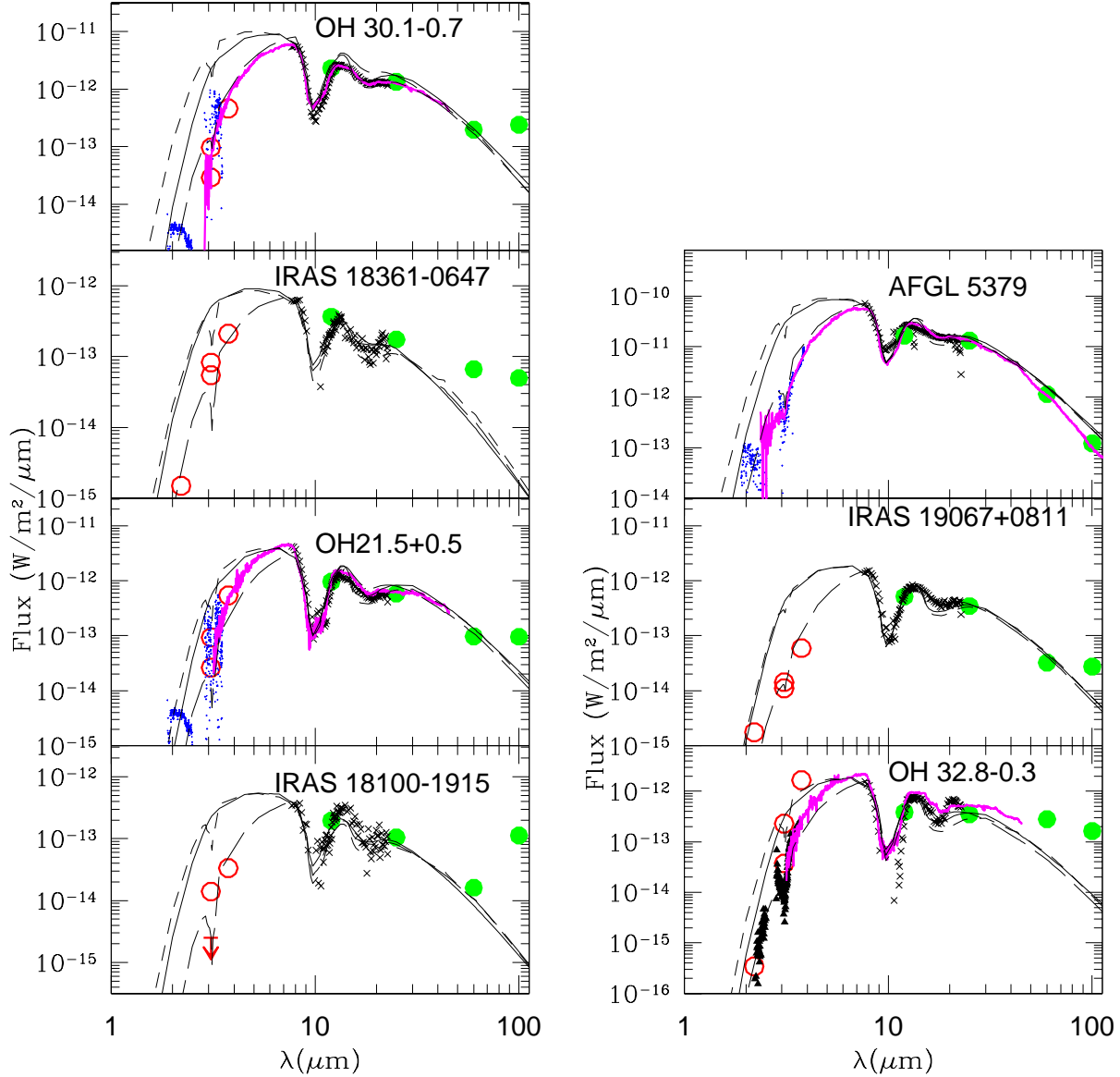}
\caption{SED fit to the stars which show ice absorption.
The symbols are the same as in Fig.~\ref{fig-sed2}. These stars
have much lower than expected near-IR flux and needs a modification
to the dust grains to fit the SED (long-dashed line). The near-IR spectrum
of OH~32.8+0.3 is taken from Justtanont \& Tielens (\cite{kay92}).
}
\label{fig-sed3}
\end{figure*}

\begin{table*}
\centering
\caption{Input parameters for the modelling and the derived silicate 
and water-ice optical depths.}
\begin{tabular}{lcccccc}
\hline\hline
    & $v$ & D & $\dot{M}_{d}$ & $\tau_{9.7}$  & $\dot{M}_{dyn}$  &$\tau_{3.1}$ \\
    & (km s$^{-1}$) & (kpc) & (M$_{\odot}$ yr$^{-1}$) & &(M$_{\odot}$ yr$^{-1}$) &\\
\hline
IRAS~18052 & 28.5$^{a}$ & 4.33 & 2.2E-6 & 18.26 & 1.1E-4 & 3.03 \\
IRAS~18092 & 17.0$^{a}$ & 6.10 & 9.3E-7 & 12.35 & 1.1E-4 & 2.05 \\
IRAS~18100 & 13.6$^{a}$ & 7.50 & 1.5E-6 & 17.71 & 2.1E-4 & 4.01 \\
OH~16.1    & 21.0$^{a}$ & 1.90 & 3.5E-6 & 32.40 & 2.9E-4 & 5.37 \\
IRAS~18257 & 18.2$^{a}$ & 6.00 & 1.4E-6 & 16.18 & 1.5E-4 & 2.68 \\
OH~21.5    & 18.1$^{a}$ & 2.50 & 2.8E-6 & 29.71 & 2.6E-4 & 6.14 \\
IRAS~18327 & 19.3$^{a}$ & 3.00 & 1.2E-6 & 13.48 & 1.2E-4 & 2.72 \\
OH~26.5    & 15.0$^{a}$ & 1.37 & 2.3E-6 & 23.55 & 2.6E-4 & 4.27 \\
IRAS~18361 & 15.9$^{a}$ & 6.00 & 1.4E-6 & 15.71 & 1.7E-4 & 3.69 \\
OH~30.7    & 17.0$^{a}$ & 3.70 & 2.3E-6 & 27.31 & 2.6E-4 & 4.99 \\
OH~30.1    & 18.1$^{a}$ & 1.75 & 2.2E-6 & 18.31 & 2.2E-4 & 4.08 \\
IRAS~18488 & 19.3$^{a}$ & 3.90 & 4.1E-6 & 37.19 & 3.6E-4 & 7.72 \\
OH~32.8    & 17.7$^{a}$ & 4.30 & 2.8E-6 & 24.84 & 3.1E-4 & 5.50 \\
IRAS~18588 & 21.5$^{a}$ & 5.30 & 1.4E-6 & 14.01 & 1.3E-4 & 2.32 \\
OH~39.9    & 14.7$^{a}$ & 5.00 & 1.1E-6 & 17.07 & 1.6E-4 & 2.83 \\
IRAS~19067 & 16.3$^{b}$ & 3.80 & 2.6E-6 & 27.89 & 2.7E-4 & 5.09 \\
IRAS~22177 & 18.3$^{c}$ & 2.10 & 1.2E-6 & 14.72 & 1.3E-4 & 2.87 \\
AFGL~5379  & 18.3$^{a}$ & 0.58 & 2.0E-6 & 18.58 & 2.0E-4 & 3.93 \\
OH~13.1    & 14.5$^{d}$ & 6.60 & 1.8E-6 & 25.86 & 2.4E-4 & 4.29 \\
IRAS~18310 & 15.1$^{b}$ & 6.60 & 1.8E-6 & 24.24 & 2.4E-4 & 4.02 \\
OH~51.8    & 18.6$^{b}$ & 5.00 & 1.7E-6 & 19.35 & 1.8E-4 & 3.21 \\
OH~65.7    & 10.8$^{b}$ & 5.00 & 1.8E-6 & 30.90 & 3.2E-4 & 5.12 \\
OH~75.3    & 11.6$^{b}$ & 5.00 & 1.8E-6 & 30.03 & 3.0E-4 & 4.98 \\
\hline
\end{tabular}
\\
$^{a}$ Sevenster (\cite{seven})\\
$^{b}$ David et al. (\cite{david})\\
$^{c}$ Kemper et al. (\cite{ciska})\\
$^{d}$ te Lintel Hekkert (\cite{hekkert})
\label{tab-mdot}
\end{table*}

Figs.~\ref{fig-sed1}, \ref{fig-sed2} and \ref{fig-sed3}
show the model fits to individual objects. Half of our sample 
of extreme OH/IR stars do not show convincing evidence for the water-ice 
absorption, despite the fact that the mass loss rate is very high. The derived
mass loss rate and other parameters are listed in Table~\ref{tab-mdot}.
The range of the dust mass loss rate of these extreme OH/IR stars are
between (1-4)\,10$^{-6}$ M$_{\odot}$ yr$^{-1}$. In deriving the dust
mass loss rate, we assume a constant dust velocity for all the grain
sizes. Although this is the case for the smallest dust grains, the drift
velocity for the largest size (0.25$\mu$m in our model) can be up to
3 km s$^{-1}$. We address this issue of the dust velocity and how it
affects the mass loss rate estimate in the next section.

\subsection{The dynamical mass loss rate}

In order to determine the total mass loss rate from a star, we used the
derived dust mass loss rate and assume that the radiation pressure on
the dust grains is the main driving force of the circumstellar wind. 
The dust momentum is transferred to the gas and knowing the terminal
velocity of the gas for each star, we can calculate the dynamical
mass loss rate (Goldreich \& Scoville \cite{goldreich}). The only unknown
in solving the equation of motion is the mass of the star. We assume
that the progenitor stars of these extreme OH/IR stars are more massive
than the hot bottom burning limit of 4 M$_{\odot}$, otherwise they
would end up as C-rich AGB stars. We assigned an arbitrary mass of 5
M$_{\odot}$ for all our stars. However, this parameter has a very small 
effect on the derived mass loss rate. A change from a 1 to 5 M$_{\odot}$
gravitational potential resulted in a decrease in the derived mass
loss rate of $\leq$ 30\%. We also calculated the dust velocity and
fed it back to the calculation for the SED hence the value of the dust
mass loss rate in Table~\ref{tab-mdot} is consistent with the
calculated dust velocity for the largest grain size.

The main uncertainty in the derived dust and dynamical mass loss rates
lie in the estimate of the distance to the star. In many cases, this is
not known and the distance was estimated by assuming the stellar
luminosity of 10$^{4}$ L$_{\odot}$. 
It was shown that semiregular variables have a mean luminosity
of 4\,200L$_{\odot}$ (Olofsson et al. \cite{hans02}) and OH/IR stars
are the more luminous counter part of these stars. An error in
the distance estimate reflects directly the error in the mass loss rate.
Fortunately, there is a direct scale
between the mass loss rate and the distance, i.e., $\dot{M}_{d}$/D is
constant in order to fit the same spectrum.
Our derived total mass loss rates for these objects are very high,
a few 10$^{-4}$ M$_{\odot}$ yr$^{-1}$, i.e., they are in a superwind phase.
Such a high rate can be sustained by the high-mass end of the intermediate
mass stars for only a relatively short period ($\sim 10^{4}$ yr).
From our calculation, the gas-to-dust mass ratio ranges from 
50 to 180, with a mean value of 110.
This is in agreement with the value of 160 from Knapp (\cite{knapp}).

%Many of the extreme OH/IR stars have their
%gas mass loss rate estimated by modelling the CO rotational lines. However,
%the rate derived are relatively low, i.e., a few 10$^{-6}$ M$_{\odot}$ yr$^{-1}$
%(Heske et al. \cite{heske}) which would imply a gas-to-dust mass ratio of 
%around unity. This seems implausible, compared to the estimated ratio of
%160 from observations of AGB stars (Knapp \cite{knapp}). The more likely explanation
%for the low gas mass loss rate, compared to that of the dust is the mass loss 
%rate of these stars have recently increased, i.e., it has entered into a 
%superwind phase. The slow expansion velocity of this superwind has not reached
%the outer part of the envelope where the low rotational CO lines originates.
%From our calculation, these stars
%are losing mass at a very high rate of a few 10$^{-4}$ M$_{\odot}$ yr$^{-1}$.

\subsection{Correlation with the optical depths}

From fitting the SED to the observations, we divide our sample into
two main categories. The first group contains stars which do not show the ice 
absorption at 3.1$\mu$m (Fig.~\ref{fig-sed1}). We modelled these stars
using pure silicate grains. The second group of stars do show ice absorption
from SIRCA and/or CGS4 observations. Their SEDs can be described by the
two components of dust, i.e., warm inner silicate grains and composite 
grains where the temperature drops below 100~K. This group can be further
subdivided into those which are well fitted by the same silicate grains
used for group 1 (Fig.~\ref{fig-sed2}), and those which show near-IR
deficiency compared to the model (Fig.~\ref{fig-sed3}). For simplicity,
we refer to groups 1, 2a and 2b. Due to the lack of near-IR spectrum of 
the stars in group 2a, we cannot firmly conclude whether the silicate dust
for these stars should be different from group 2b. We will discuss
this point on the following section.
%, i.e., the distinction
%between groups 2a and 2b may simply be due to the deeper water-ice
%absorption in the latter.

\begin{figure}
\resizebox{\hsize}{!}{\includegraphics{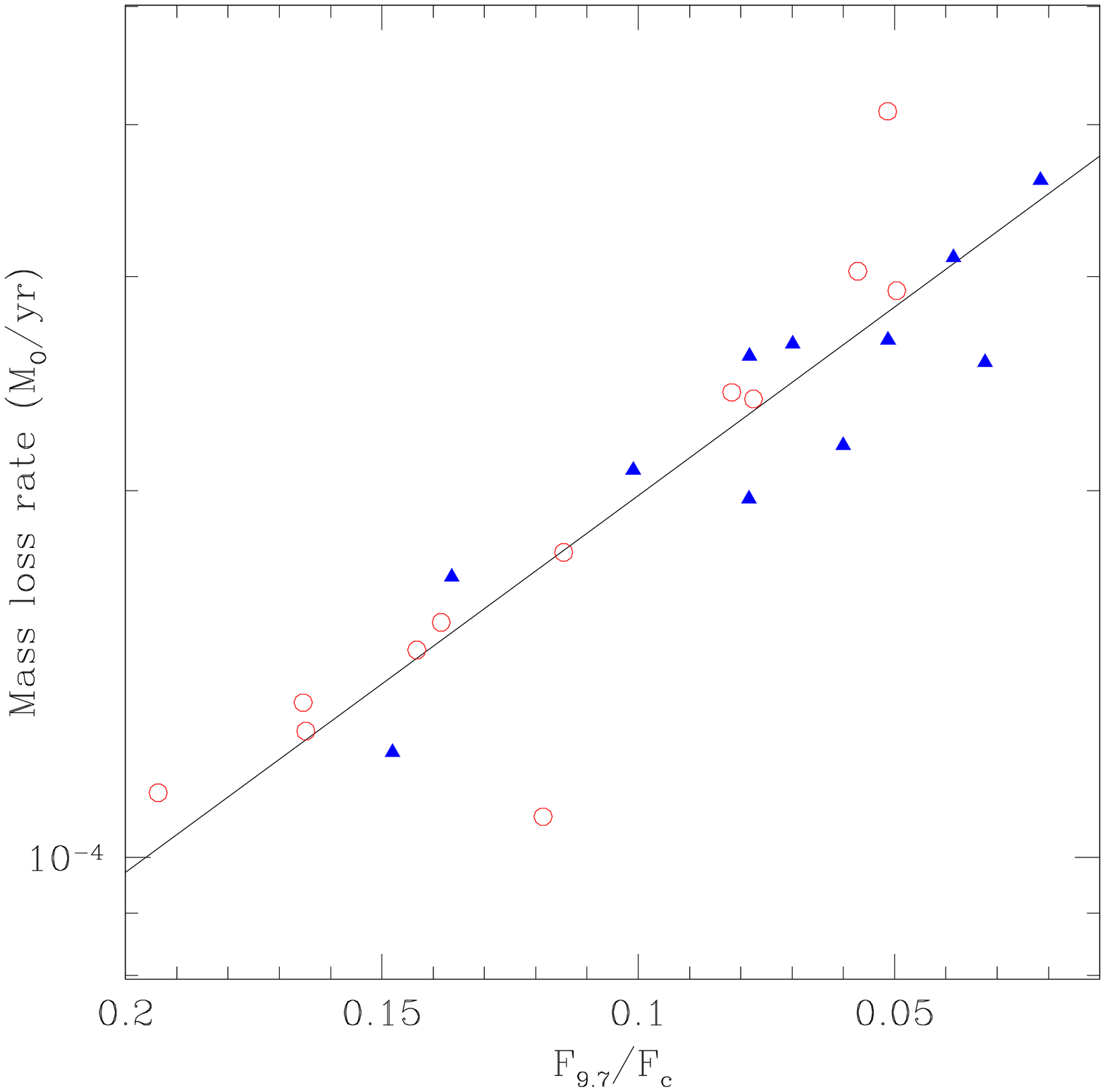}}
\caption{A plot of the derived mass loss rates against the
depth of the 9.7$\mu$m silicate absorption. The line represents
the best fit to the data. The filled and opened symbols are stars with
and without water-ice absorption, respectively.
}
\label{mdot_sildepth}
\end{figure}

\begin{figure}
\resizebox{\hsize}{!}{\includegraphics{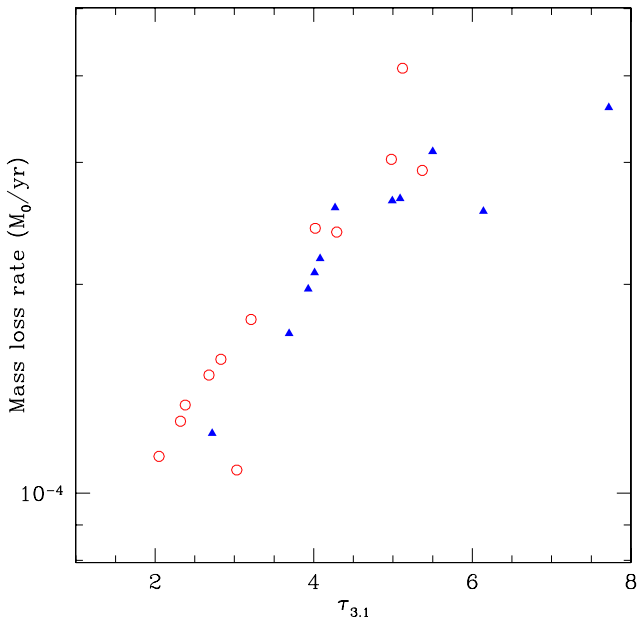}}
\caption{A plot of the mass loss rates against the optical 
depths at 3.1$\mu$m for stars with (solid triangle) and
with no (open circle) detected water-ice absorption band. 
}
\label{fig-mdot_tau}
\end{figure}

We find a strong correlation between the derived mass loss rate and
the depth of the  9.7$\mu$m feature (Fig~\ref{mdot_sildepth}). 
We assume here that the 
interstellar contribution is negligible compared to the circumstellar 
contribution. The dependence
is not surprising as the depth of the silicate feature 
is a direct measure of the amount of dust along the line of sight. 
The relationship between the measurable
depth, simply the ratio of the flux at the bottom of the silicate feature and 
the continuum over the feature at the same wavelength can be 
expressed as
\begin{equation}
 log~\dot{M} = -3.1~ \frac{F_{9.7}}{F_{\rm cont}} ~ -~ 3.4
\end{equation}
Here, we defined a continuum flux at 9.7$\mu$m as a linear interpolation 
of the fluxes between wavelength of 8 and 12$\mu$m over the silicate 
absorption feature.
The uncertainty of the relationship is a factor of 0.2 on the log 
scale (3$\sigma$) which translates to a factor of 1.6 in the mass loss rate.
A similar relationship between mass loss rate and the strength of
the silicate feature 
has been shown to be the case for optically thin AGB 
stars and supergiants using the silicate emission (Skinner \& Whitmore 
\cite{skinner}). 
Our derived relation simply extends the estimation of the mass loss rate
into the very optically thick regime, i.e., $\dot{M} \geq
10^{-4}$ M$_{\odot}$ yr$^{-1}$.

In Fig.~\ref{fig-mdot_tau}, we show the plot of the derived
mass loss rate and the derived optical depth at 3.1$\mu$m, $\tau_{3.1}$
from the SED fitting. 
Note that the value of the optical depth in Table~\ref{tab-mdot}
is the total optical depth due to both the pure silicate grains and silicate
core-ice mantle grains hence the values are relatively high.
The correlation is very strong, regardless of whether
the star shows the ice absorption band. We find the same fit
even if the stars in group 1 (those with no water-ice) are not taken 
into account. This confirms
the correlation seen earlier by Meyer et al. (\cite{meyer}) for a smaller
sample of star. They stipulated a direct trend that a star with a high mass
loss rate shows deeper ice absorption and hence higher column density
of water-ice. It is somewhat
interesting that stars in group 1 also show the same trend of higher $\tau_{3.1}$
with the mass loss rate. This reflects the very high silicate dust absorption
efficiency without invoking other dust type.
Note here that the optical depth in
table~\ref{tab-mdot} and Fig.~\ref{fig-mdot_tau} is the total
optical depth integrated along the line-of-sight and includes the
contribution of both silicate and water-ice. Most of the opacity
is due to the warm silicate dust which rises with decreasing wavelength
hence giving the correlation seen for group 1 stars.
As can be seen in Fig.~\ref{fig-mdot_tau}, not all stars
with very thick circumstellar shell show evidence of ice formation. 
There is also a slight hint that the relationship may taper off towards
higher optical depth and mass loss rate.

\subsection{Ice condensation}

It seems that even though the condition for ice formation is a highly optically
thick envelope (high density) so water-ice can be shielded from the
central stellar radiation field (low temperature),
not all the stars in our
sample show the ice absorption band at 3.1$\mu$m. From simply knowing the 
mass loss rate, it is not possible to predict which star has the icy
mantle grains. However, it is true that for those which have water-ice 
signature, the mass loss rate correlates with the optical depth, and hence the
ice column density, $N_{ice}$, at 3.1$\mu$m via 
(d'Hendecourt \& Allamandola \cite{louis})
\begin{equation}
       N_{ice} = \tau_{3.1} \nu / {\it A}
\end{equation}
where $\nu$ and {\it A} are the full-width-half maximum and the absorbance
of the water-ice band at 3.1$\mu$m, respectively.

It is a puzzle why some stars form water-ice and not others. One possible
explanation is the water vapour abundance in the photosphere and circumstellar
envelope. We therefore investigate the existing spectra of these stars.
It is unfortunate that the near-IR spectra of the sample stars
are not good enough to see the stretching bands of water at 3$\mu$m 
from the available ISO-SWS data as the signal-to-noise ratio in these 
spectra is very low.
The bending band around 6$\mu$m were then checked to see
whether there are corresponding lines. In this region, there
are expected to be strong lines which are sensitive to the temperature of
the gas, i.e., at 6.59, 6.63 and 6.83$\mu$m.
\begin{figure}
\resizebox{\hsize}{!}{\includegraphics{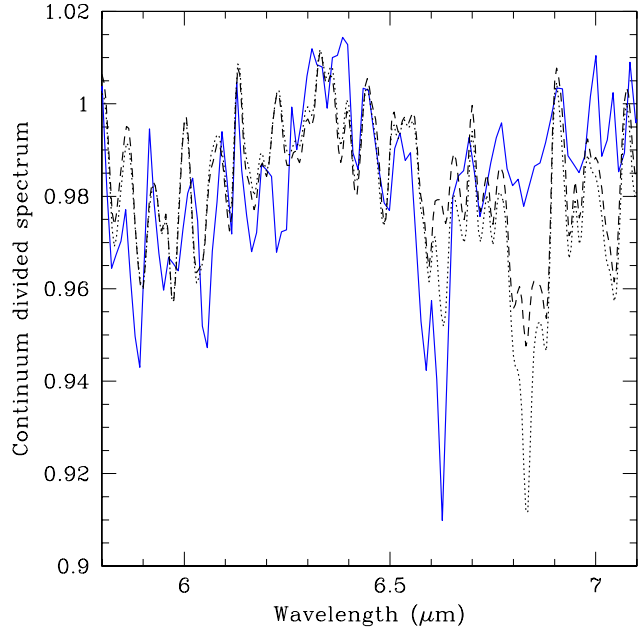}}
\caption{The ISO-SWS continuum divided spectrum of OH26.5+0.6 (solid line)
compared with the calculated absorption spectrum for the gas temperature
of 300K (dashed line) and 500K (dotted line).
}
\label{fig-condiv}
\end{figure} 
The ratio of the first two lines are a good temperature indicator
(Fig.~\ref{fig-condiv}).These lines are seen in the continuum divided spectra
of the sample. The ratio indicates a temperature in excess of 500K. However,
the expected strong feature at 6.83$\mu$m is not distinctly detected,
pointing to a lower (T$\leq$300K) temperature gas.
Other weaker features cannot be confirmed
so the detection of gaseous water remains tentative. The overlap
between our sample and the SWS spectra happens to be those stars which show
water-ice signature. The lack of the detection of water vapour lines may not 
be surprising taking into account that much of the available water
freezes out onto grains. In this scenario, gaseous water does not provide
a significant opacity source in the near-IR.

In order for water-ice to condense, the density has to be high and the
grains must be cool enough for the water to condense onto the
existing silicate grains. This requires that the dense circumstellar
envelope has to be of an order of a few 10$^{16}$ cm as the dust temperature
falls off as $r^{-0.4}$, according to our radiative transfer
calculation. This is in accord with the calculation by Dijkstra et al.
(\cite{rien2}) who estimated the radius of the water-ice condensation
zone started from around (1-2)~10$^{3}$ AU, the exact distance depending
on dust grain radius. 
However, we are not able to measure
the radius of the superwind and it is possible that in some of the stars,
the onset of the superwind is very recent that the outer edge of the
shell has no time to cool down below the water-ice condensation temperature.
In such a case, we do not expect to detect the 3.1$\mu$m absorption.

\section{Near-IR opacity in icy envelopes}

It is clear that in stars with a clear signature of water-ice absorption
in the envelope (group 2b), the near-IR is suppressed compared to those with a small
amount of (group 2a), or no (group 1), water-ice. The spectrum shortward
of 5$\mu$m cannot be explained by the presence of the same type of
silicate or the water-ice. In fitting the spectrum of these stars,
an adhoc near-IR opacity is included in the silicate grains 
(see Fig~\ref{fig-sed3}). As we did not see a clear evidence for the
gaseous water in the spectra of these objects, we search for an
explanation in terms of dusty components to provide the needed
opacity.

It is known that very optically thick envelopes around extreme OH/IR stars
exhibit crystalline silicate feature in the far-IR (e.g., Cami et al., 
\cite{cami}). However, the near-IR opacity of 
the crystalline silicate is much lower than that of the amorphous grains 
(Koike et al., \cite{koike}). This option cannot then be used to
explain the near-IR deficiency seen.

Kemper et al. (\cite{kemper}) reported on an extreme OH/IR star, OH127.8+0.0,
which also has a near-IR deficiency. Although this star show an evidence 
for water-ice in its spectrum, simply including it in the modelling could
not explain the observed deficiency.
They, instead, proposed metallic iron as being the absorber in order to
fit the observed spectrum. It remains curious that the lack of near-IR 
flux is seen to be accompanied by the water-ice absorption,
as  is the case in our sample stars.
It is possible
that when the water-ice condensed onto the silicate grains, some
other materials are incorporated in the matrix leading to the
extra absorption in the near-IR. 
%Somehow the amorphous silicate  which has a mixture of both Mg
%and Fe has been processed in such a way to reduce the chemical abundance
%of Fe from the grain. It is then viable that the excess Fe condenses
%in the outflow giving a higher near-IR opacity in such stars. What is
%unclear is the link between Fe and water-ice condensation.
However, the link between the condensation of water-ice and
iron grains is unclear, if the extra opacity in the near-IR is due to
iron dust. Thermodynamically, the two are not linked due to the 
difference in the condensation temperatures. Iron is expected 
condense close to the central star (Gail \& Sedlmayr \cite{gail}).
By removing iron from the gas phase, it naturally leads to the
explanation of magnesium-rich silicate grains. 
Many OH/IR stars have been found to have the signature of fosterite, 
a pure Mg-silicate (Waters \& Molster \cite{rens}).
Although it is
viable that the near-IR opacity is due to iron, it remains to
be seen why iron condenses out (close to the star) at the
same time water-ice condenses onto cold silicate grains farther out.

Alternatively, this could simply be due to the way these objects are
viewed. It is possible that these stars have a disk which is where
the water-ice resides. When this disk is along the line of sight (edge-on 
disk), the absorption is at its maximum and hence the the near-IR radiation is
more efficiently absorbed by the disk while stars which the disk
is perpendicular to the line-of-sight will show much less absorption.
Such models have been demonstrated by Ueta \& Meixner (\cite{ueta}) who
calculated a 2-D dust radiative transfer model. When viewed edge-on,
the near-IR flux drops relative to when the disk is viewed face-on.
If this is the case,
it implies that already at this stage of the AGB evolution, the
star can be seen to depart from having spherically symmetric outflow.
The presence of a disk will enhance further mass loss in an axisymmetric
manner. It is known that most of the observed post-AGB 
stars show axisymmetric morphology (Ueta et al. \cite{ueta2}).

\section{Enhanced mass loss - superwind}

\begin{table}
\caption{Calculated gas mass loss rates from dynamical calculation
(present work) and those derived from OH luminosity and CO observations.}
\begin{tabular}{lccc}
\hline\hline
Star      & this work   & OH & CO   \\
\hline
IRAS~18052 & 1.1E-4 & 6.3E-05$^{a}$ & - \\
IRAS~18092 & 1.1E-4 & 2.0E-05$^{a}$ & - \\
IRAS~18100 & 2.1E-4 & 4.0E-05$^{a}$ & - \\
OH~16.1    & 2.9E-4 & 4.7E-05$^{a}$ & - \\
IRAS~18257 & 1.5E-4 & 9.4E-05$^{a}$ & - \\
OH~21.5    & 2.6E-4 & 8.1E-05$^{a}$ & - \\
IRAS~18327 & 1.2E-4 & 5.0E-05$^{a}$ & - \\
OH~26.5    & 2.6E-4 & 9.0E-05$^{a}$ & 8.0E-6$^{d}$ \\
IRAS~18361 & 1.7E-4 & 5.1E-05$^{a}$ & - \\
OH~30.7    & 2.6E-4 & 4.5E-05$^{a}$ & - \\
OH~30.1    & 2.2E-4 & 7.5E-05$^{a}$ & 1.2E-5$^{e}$ \\
IRAS~18488 & 3.6E-4 & 6.5E-05$^{a}$ & - \\
OH~32.8    & 3.1E-4 & 1.6E-04$^{b}$ & 3.1E-5$^{e}$ \\
IRAS~18588 & 1.3E-4 & 2.5E-05$^{a}$ & - \\
OH~39.9    & 1.6E-4 & 6.6E-05$^{a}$ & - \\
IRAS~19067 & 2.7E-4 & 4.9E-05$^{a}$ & - \\
IRAS~22177 & 1.3E-4 & 1.0E-04$^{b}$ & 6.5E-5$^{d}$ \\
AFGL~5379  & 2.0E-4 & 5.3E-05$^{a}$ & 3.4E-6$^{f}$ \\
OH~13.1    & 2.4E-4 & 8.3E-05$^{b}$ & - \\
IRAS~18310 & 2.4E-4 & 3.8E-05$^{c}$ & - \\
OH~51.8    & 1.8E-4 & 8.8E-05$^{c}$ & - \\
OH~65.7    & 3.2E-4 & 2.6E-05$^{c}$ & - \\
OH~75.3    & 3.0E-4 & 5.9E-05$^{c}$ & - \\
\hline
\end{tabular}
\\
$^{a}$ Sevenster (\cite{seven})\\
$^{b}$ te Lintel Hekkert et al. (\cite{hekkert})\\
$^{c}$ David et al. (\cite{david})\\
$^{d}$ J=3-2 from Kemper et al. (\cite{ciska})\\
$^{e}$ J=2-1 from Heske et al. (\cite{heske})\\
$^{f}$ J=2-1 from Loup et al. (\cite{loup})\\
\label{tab-ohco}
\end{table}
To investigate further, we check the OH 1612~MHz maser flux for each object.
Here, there is no real correlation between the 
maser luminosity and the ice condensation as the stars with and without
ice are intermingled. It also does not follow the expected trend between
the OH maser luminosity and the mass loss rate, namely $\dot{M}/v \propto$ 
L$_{OH}^{0.5}$ (Baud \& Habing \cite{baud}). 
For an expected OH luminosity, the derived dust mass loss rate is much
higher than expected. This deviation can be 
understood in the context of the superwind. These stars have most likely
recently undergone an increase in the mass loss rate which has not have time
to propagate out to where the OH maser emission originates from. Typically,
OH maser scale is of an order of 10$^{17}$cm which reflects the mass loss
rate of about 2\,000 years ago. The discrepancy between the dynamical
and the expected mass loss rate derived from the OH luminosity reflects the 
time scale of the superwind of less than about (1-2)\,10$^{3}$ years old.
An enhanced mass loss rate varies by about a factor of 2-6 compared to 
those derived from the OH maser. 

This is further confirmed when looking at available CO observations of
our sample. Only very few have been observed and detected 
(Table~\ref{tab-ohco}).
Their CO mass loss rate is significantly lower than the derived total
dynamical mass loss rate by a factor of 10 or more. The estimated
CO mass loss rate is based on the formalism by Olofsson et al. 
(\cite{hans93})
and Gronewegen et al. (\cite{martin}) who extended the equation for CO 
J=3-2. 
\begin{equation}
\dot{M} = \frac{1.4~T_{\rm MB} v^{2} D^{2} \theta^{2}}{2\,10^{19} f_{\rm CO} s(J)}
\end{equation}
where $T_{\rm MB}$ is the main beam temperature, $v$ is the terminal velocity
in km s$^{-1}$, D is the distance in pc, $\theta$ is the FWHM of the beam,
$f_{\rm CO}$ is the CO abundance (taken to be 5\,10$^{-4}$ here) 
and s(J) is a constant, depending on the rotational transition.
The CO mass loss rate can be seen to be lower than the OH
mass loss rate which supports that CO transitions probe the much
earlier mass loss episode than that of OH and the infrared. 
The disparity between the dynamical mass loss rate and those derived
from CO observations lends a very strong support to the superwind idea.

By comparing the OH mass loss rate and the dynamical mass loss rate
(Table~\ref{tab-ohco}), we can determine the upper limit of the 
lifetime of the superwind of 2\,000 yrs. This gives enough time for 
the dust to cool below the water-ice condensation temperature.
What is not clear is the actual time since the superwind started in
each object. In some stars, the superwind might have just started
a few hundred years ago hence grains are too hot for water condensation.
However, Justtanont et al. (\cite{kay96}) showed that OH26.5+0.6
entered the superwind phase some 200 years ago and still the spectrum
shows the evidence for water-ice (Meyer et al \cite{meyer}). This means
that the high density of the superwind can effectively shield the stellar 
radiation, allowing silicate grains to be cool enough for water-ice 
condensation.
In order to determine the extent of a superwind, high resolution
observations in the infrared, or CO and modelling will be
needed (e.g., Fong et al., \cite{fong}).
It may also be possible to probe the superwind using high
angular resolution OH maser observations as these stars are
known to have relatively strong maser emission.

\section{Summary}

We search the circumstellar envelope of extreme OH/IR stars 
for the presence of water-ice in an attempt to check the
reported correlation between the water-ice column density
and the mass loss rate (Meyer et al \cite{meyer}). We found
that such a correlation exists in our sample stars which display
the 3.1$\mu$m water-ice absorption. However, not all the stars
in our study show the water-ice band. Furthermore, from the
depth of the silicate 9.7$\mu$m band which is an indication of the
dust mass loss rate, it is not possible to predict which of
the circumstellar shells should display the water-ice band. Stars which show 
a distinct water-ice absorption band also shows a near-IR deficiency
compared to stars with no water-ice detection. A plausible 
explanation is another dust species which condensed at the same time
as the water-ice, e.g., iron (Kemper et al. \cite{kemper}). 
It remains a puzzle how the two species with different 
condensation temperatures are linked.
It is also possible that the water-ice is present in a disk
around the central star and when viewed along the line-of-sight,
gives higher extinction than at an inclined angle.

We found a simple estimate of the mass loss rate from
the ratio of the depth of the silicate absorption feature at 9.7$\mu$m
and the continuum flux. The main uncertainty in our estimated
dust mass loss rate is the distance to the stars which in
many of the cases here is unknown.

Comparison of our derived dust mass loss rate with the expected
gas mass loss rate from the OH luminosity and CO rotational observations
indicates that these 
extreme OH/IR stars have recently undergone an increase in the
mass loss rate by a factor of $\geq$ 10. The maximum
timescale for the superwind is less than the time it takes 
for the mass loss to reach the H$_{2}$O photodissociation radius, 
i.e., $\sim$ 2\,000 yrs. In some cases, this can be much smaller
%and hence explain the lack of water-ice in some circumstellar
%envelopes of extreme OH/IR stars.
as the dynamical mass loss rate is higher than the mass loss rate
derived from the OH observations. High angular resolution observations
($\leq 0.1^{\prime\prime}$) will be able to confirm the extent
of the superwind.

\end{document}